\documentclass[runningheads]{llncs}
\usepackage[T1]{fontenc}
\usepackage{graphicx}
\usepackage{amsmath}
\usepackage{amssymb}
\DeclareMathOperator*{\argmax}{arg\,max}
\begin{document}
\title{An Efficient and Sybil Attack Resistant Voting Mechanism}
\author{Jeremias Lenzi}
\institute{Department of Business and Economics, University of Basel}
\maketitle              
\begin{abstract}
Voting mechanisms are widely accepted and used methods for decentralized decision-making. Ensuring the acceptance or legitimacy of the voting mechanism's outcome is a crucial characteristic of any robust voting system. Consider the following scenario: A group of individuals wants to choose an option from a set of alternatives without requiring an identification or proof-of-personhood system. Moreover, they want to implement utilitarianism as their selection criteria. In such a case, players could submit votes multiple times using dummy accounts, commonly known as a Sybil attack (SA), which presents a challenge for decentralized organizations. To be functional and sustainable, they need to address this issue without harming their selection criteria. Is there a voting mechanism that always prevents players from benefiting by casting votes multiple times (SA-proof) while also selecting the alternative that maximizes the added valuations of all players (efficient)? One-person-one-vote is neither SA-proof nor efficient. Coin voting is SA-proof but not efficient. Quadratic voting is efficient but not SA-proof. This study uses Bayesian mechanism design to propose a solution. The mechanism's structure is as follows: Players make wealth deposits to indicate the strength of their preference for each alternative. Each player then receives a non-negative amount based on their deposit and the voting outcome. The proposed mechanism relies on two main concepts: 1) Transfers are influenced by the outcome in a way that each player’s optimal action depends only on individual preferences and the number of alternatives; 2) A player who votes through multiple accounts slightly reduces the expected utility of all players more than the individual benefit gained. This study demonstrates that if players are risk-neutral and each player has private information about their preferences and beliefs, then the mechanism is SA-proof and efficient. This research provides new insights into the design of more robust decentralized decision-making mechanisms.
\end{abstract}
\section{Introduction}
Suppose a collective desires to implement a decentralized decision-making procedure to select an option from a set of alternatives whenever a choice problem is faced. Moreover, let's assume that the collective wants to implement an efficient mechanism, so the procedure should always select the alternative that maximizes added valuations of players independently from each particular choice problem (a utilitarian outcome), given that each member possesses private information regarding their preferences and beliefs. Also, let's assume that the collective wants to allow players to vote multiple times since it doesn't want to rely on identification systems or proof-of-personhood mechanisms.   

Voting mechanisms are accepted and used procedures to meet different pre-accorded selection criteria as classic democratic theory states, Dahl (2008) ~\cite{ref_book1}. A selection criterion is an accepted rule by all members of a collective. If the alternative selected by the mechanism meets the selection criterion rule, all members will agree with the mechanism's outcome. An implementable voting mechanism should always generate an outcome such that the selected alternative always meets the selection criterion. Otherwise, the decision made by the procedure might face legitimacy issues and could lead to sub-groups of agents not accepting the decision made. For example, a standard one-person-one-vote voting mechanism (1P1V) always selects the alternative that is of preference to more individuals. Notice that this selection criterion doesn't account for the intensities of preferences. The 1P1V mechanism requires aid from a system that ensures that each agent votes only once. In a scenario where players could cast multiple votes, the selected alternative might not be the one that mmets the selection criteria \footnote[1]{This is why we don't see 1P1V implemented in blockchain environments since players could vote from multiple addresses}. Submitting multiple votes is usually referred to as a Sybil attack.    

As a solution to this problem, we propose a voting mechanism based on transfers. Players will deposit money in \textit{abstract envelopes} that supports the mechanism with all the information it needs to generate an outcome that meets the selection criteria at equilibrium. After an alternative is selected, players will receive back an envelope with money gathered from the deposits made by all players. The amount of money they receive back follows a known rule.

The selection criteria that this work focuses on is the utilitarian rule. This rule states that the chosen alternative must be the one that maximizes the added valuations of players over the set of options.
Mechanism design applied to voting mechanisms that implement the utilitarian rule can be broadly found in the literature such as VCG mechanisms of Vickrey (1961), Clarke (1971) and Groves (1973)~\cite{ref_article1,ref_article2,ref_article3} and the AGV mechanisms by d’Aspremont and Gerard-Varet (1979) and Arrow (1979)~\cite{ref_article4,ref_article5}. Other examples are Quadratic Voting mechanisms ~\cite{ref_article6,ref_article7,ref_article8,ref_article9} on which we base our approach. We use a similar design that pushes players to have a concave quadratic utility function. However, we obtain the negative quadratic term by using a different concept. Instead of the cost of votes being a value proportional to its square, we obtain the negative quadratic term by multiplying two linear functions, the player's expected value of transfers. 
%

Blockchain communities and decentralized autonomous organizations face several choice problems throughout their lifespan, such as updating protocols, financing projects, reversing harms from hacks, etc. Some of these types of organizations don't have or prefer not having an identification mechanism to distinguish their members (because of ideological or foundational reasons). %
One of the most implemented governance structures in blockchain environments is what is called \textit{coin voting} or \textit{holders voting}, a type of weighted voting mechanism where the weight or amount of votes of a player is equal to the holdings of that player of a particular \textit{governance token}\footnote[2]{Players have to lock their holdings while the voting process is in process, so no double spending is possible.}, Fan et al. (2023) ~\cite{ref_article11}. Holdings could be dispersed into several accounts so players could vote multiple times through different accounts. However, the total amount of holdings of the governance tokens of each player remains the same over all the accounts. Voting from these accounts would imply a total of votes equal to voting from the original account. In other words, coin voting is Sybil attack-proof. However, this mechanism selects the alternative that gathers more holdings, and it doesn't take into account players' intensity of preferences. Its selection criterion differs from the one this work focuses on.


This work studies a voting mechanism over a risk-neutral population. It shows that the mechanism presents the following properties in Bayes-Nash equilibrium: 1) Non-indifferent players are always incentivized to vote. 2) The mechanism always selects the alternative that maximizes the added valuations of players. 3) Players don't benefit from voting multiple times. 4) Exists a non-negative surplus. Under the proposed mechanism, players can express an intensity of support or rejection towards any alternative by making corresponding deposits. The mechanism gathers these deposits and uses them to calculate the corresponding intensities of preference (votes) for each alternative, which will be aggregated across players to determine a selected alternative. Given this selected alternative and the corresponding deposits made by players, the mechanism will determine the amounts of the transfers to players after the voting process. The mechanism gathers all the information it needs from the inputs it receives, and no calibration or fine-tuning is needed before each particular choice problem.
\section{Model}
Let's assume a set of $n \geq 1$ risk-neutral rational expectation maximizer players that face a 
choice problem between $m \geq 1$ alternatives $A=\{A_1,...,A_m\}$. Each player is assigned by Nature a valuation vector $u_i = (u_{ij})_{j=1}^m$ with $u_{ij} \in [0,w]$ from a certain distribution $\mathcal{F}$ unknown by players, with $\omega \in \mathbb{R}$ common knowledge. Each player receives a private signal with their valuation vector and a prior distribution over other players' valuations. We are going to suppose a coalition-free environment.
To simplify the scope of our analysis, we restrict the family of priors' distributions. We then assume that a player's belief is such that the probability of alternative $A_j$ being selected given that she votes $x_i$ and other players play optimally is the following 
\begin{equation}
    P(A_j|x_i,x_{-i}^*) = P_{ij}^{(0)} + p_i(x_{ij} - \frac{1}{m-1}\sum_{r\neq j} x_{ir}) 
\end{equation}
with $\sum_{j=1}^m P_{ij}^{(0)} = 1$ and $p_i \in (0,\varepsilon(\omega))$. The parameter $P_{ij}^{(0)}$ is the probability of alternative $j$ being selected if player $i$ doesn't participate. The parameter $p_i$ is the marginal probability of adding votes for all alternatives. Both parameters will determine the belief of the player. The parameter $\varepsilon(\omega)$ is a small enough positive number depending on $\omega$ to ensure well-defined probabilities.

The previous assumption restricts the family of priors our model handles. We consider a restricted model since the objective of this article is to present the ideas behind the mechanism. However, the mechanism could be extended using the same concepts to incorporate a broader family of priors.

\subsection{Proposed Mechanism}
Each player grabs an envelope with $m$ small compartments, one for each alternative. To acquire votes, players will make non-negative deposits over each alternative given their preference over them. We will denote the deposits made by player $i$ as a vector $D_i = (D_1,...,D_m) \in \mathbb{R}_{\geq 0}^m$. Votes of each player for each alternative are going to be calculated given $D_i$ and will be denoted by $x_i = (x_{i1},...,x_{im}) \in \mathbb{R}^m$. Notice that we allow votes to be negative. After the mechanism gathers deposits from all players, it calculates votes' values and adds votes for each alternative. Then, the mechanism selects the alternative that adds more votes. Finally, it calculates the amounts $R_i$ that will be transferred to each player from the collected deposits.

Votes $x_{ij}$ for each alternative $j$ are calculated in the following way given the deposit vector of a player $D_i$,

\begin{equation}
    x_{ij} = \frac{1}{a(m-1)} (D_{ij} - t_{ij}) \quad , \quad t_{ij} = \frac{1}{m-1}\sum_{r\neq j} D_{ir}
\end{equation}
where $a = (\frac{3}{2})^{h}$ is a parameter carefully selected since it will play an important role in incentivizing players to participate only once. $a$ depends on $h$, which denotes the number of participants. A player could participate multiple times by delivering more than one envelope, so the number of participants $h$ could differ from the number of players $n$. 
The parameter $t_{ij}$ has a particular objective, it guarantees $\sum_{j=1}^m x_{ij} = 0$ for any deposit vector $D_i$. Having the term $t_{ij}$ in the mechanism allows a relativistic approach, measuring the intensity of preference relative to the mean of valuations of all alternatives.

After votes values are calculated, the mechanism adds them for each alternative over all participants and selects the alternative that adds more votes.
\begin{equation}
    \argmax_{j} \quad \sum_{i=1}^h x_{ij}
\end{equation}

The transfer $R_i$ that each participant will receive back will depend on the selected alternative.
\begin{equation}
    R_i = \sum_{j=1}^m (R_{ij}^{(0)}+R_{ij}^{(1)}+t_{ij})
\end{equation}
where $R_{ij}^{(0)}$ and $R_{ij}^{(1)}$ are calculated given the following criteria.
If adding the votes for alternative $j$ by replacing the votes of player $i$ for all alternatives to zero would imply that alternative $j$ is not selected, then $R_{ij}^{(0)} = a \sum_{r\neq j}x_{ir}$. Otherwise, if $j$ is the selected alternative then $R_{ij}^{(0)} = 0$.
If adding the votes for alternative $j$ only taking into account the votes of player $i$ for alternative $j$ and replacing the votes of this player for the rest of the alternatives to zero, would imply that alternative $j$ is not selected, then $R_{ij}^{(1)} = a x_{ij}$. Otherwise, if $j$ is the selected alternative then $R_{ij}^{(1)} = 0$.

Consider an illustrative example where $h=2$ and $m=3$. Suppose the deposit vectors of each participant are such that $x_{1} = (3,1,-4)$ and $x_{2} = (-3,2,1)$. Then, by adding the votes for each alternative 
\begin{equation*}
    x_{11} + x_{21} = 0 \quad , \quad x_{12} + x_{22} = 3 \quad , \quad x_{13} + x_{23} = -3 
\end{equation*}
we have that the mechanism selects the alternative $A_2$ since it is the alternative that gathers more votes.
Let's calculate $R_i$ for participant $i=1$. Using $x_1 = (0,0,0)$ and maintaining the original $x_2 = (-3,2,1)$. Then, by adding votes for each alternative  
\begin{equation*}
    0 + x_{21} = -3 \quad , \quad 0 + x_{22} = 2 \quad , \quad 0 + x_{23} = 1
\end{equation*}
we get that $A_2$ would be selected in this virtual scenario. Then, we can calculate   
\begin{equation*}
    R_{11}^{(0)} = a(1+(-4)) = -3a  \quad , \quad R_{12}^{(0)} = 0 \quad , \quad R_{13}^{(0)} = a(3+1) = 4a
\end{equation*}
Similarly, to calculate $R_{1j}^{(1)}$ the mechanism modifies the vector $x_1$ by setting $x_{1r} = 0$ for $r \neq j$. Then, \\

    Using $x_1 = (3,0,0) \quad \longrightarrow \text{ } 3 + x_{21} = 0 \text{ } , \text{ } 0 + x_{22} = 2 \text{ } , \text{ } 0 + x_{23} = 1$\\
    
    Using $x_1 = (0,1,0) \quad \longrightarrow \text{ } 0 + x_{21} = -3 \text{ } , \text{ } 1 + x_{22} = 3 \text{ } , \text{ } 0 + x_{23} = 1$\\

    Using $x_1 = (0,0,-4) \text{ } \longrightarrow \text{ } 0 + x_{21} = -3 \text{ } , \text{ } 0 + x_{22} = 2 \text{ } , \text{ } -4 + x_{23} = -3$
\\~\\
In all these virtual scenarios $A_2$ is selected, then
\begin{equation*}
    R_{11}^{(1)} = 3a \quad , \quad R_{12}^{(1)} = 0 \quad , \quad R_{13}^{(1)} = -4a 
\end{equation*}
Let's notice that in this particular example we have
\begin{equation*}
    \sum_{j=1}^m (R_{1j}^{(0)} + R_{1j}^{(1)}) = 0
\end{equation*}
Which implies that
\begin{equation*}
    R_i = \sum_{j=1}^m t_{ij} = \sum_{j=1}^m \frac{1}{m-1} \sum_{r\neq j} D_{i,r} = \frac{1}{m-1}(m-1)\sum_{j=1}^m D_{ij} = \sum_{j=1}^m D_{ij} 
\end{equation*}
So, we can observe that participant $i=1$ receives all her deposits back, however it's not always the case.
\subsection{Equilibrium}
Players want to answer the following question: How much should I vote for each alternative given the intensities of my preferences? To answer this question, players select the voting vector that maximizes their expected value, given that other players are playing optimally.
\begin{equation}
   x_i^* := \argmax_{x_i} \quad U_i(x_i,x_{-i}^*)
\end{equation}
The utility for player $i$ given the realization of vector $u_i$ is
\begin{equation}
    U_{i}(x_i,x_{-i}^*) := \sum_{j=1}^m (u_{ij}P_{ij} - D_{ij} + \underbrace{(1-P_{ij}^{(1)})R_{ij}^{(1)} + (1-P_{ij}^{(0)})R_{ij}^{(0)} + t_{ij}}_{\mathbb{E}[R_{ij}]}) 
\end{equation}
where, 
\begin{equation*}
    P_{ij} := P(A_j|x_i,x_{-i}^*) \quad , \quad 
    P_{ij}^{(0)} := P(A_j|x_i=0,x_{-i}^*) 
\end{equation*}
\begin{equation*}
    \text{ } P_{ij}^{(1)} := P(A_j|x_{ij},x_{ir}=0,x_{-i}^*) \quad , \quad r\neq j
\end{equation*}
The utility function is composed by adding five terms over each alternative $j$. The first term $u_{ij}P_{ij}$ represents the expected value of alternative $j$. This term depends on the player's valuation for alternative $j$, player $i$'s belief, and her voting vector $x_i$. The second term is a subtraction of $D_{ij}$, which is the deposit player $i$ makes for alternative $j$. The third term represents the probability that alternative $j$ is not selected given that $x_{ir} = 0$ for $r\neq j$ multiplied by the transfer the player will receive in that scenario \footnote[3]{We perform an abuse of notation by denoting $R_{ij}^{(0)}$ and $R_{ij}^{(1)}$ as the non-zero term that variable may present.}. The fourth term is the probability that alternative $j$ is not selected given that $x_{ir} = 0$ for all $r=1,...,m$. The fifth term is the mean of deposits made by that participant for the other alternatives.

Rewriting in the utility function the explicit probabilities that the player has given her beliefs and her vote vector, we get 
\begin{equation*}
    U_i(x_i,x_{-i}^*) = \sum_{j=1}^m (u_{ij}(P_{ij} + p_i(x_{ij} - \frac{1}{m-1}\sum_{r \neq j} x_{ir})) - (a(m-1)x_{ij} + t_{ij}) 
\end{equation*}
\begin{equation}
    + (1 - (P_{ij} + p_ix_{ij}))ax_{ij} + (1-P_{ij})a\sum_{r \neq j} x_{ir} + t_{ij} ) 
\end{equation}
%
The player's problem is finding the voting vector $x_i$ that maximizes her utility function, given that other players play optimally. To obtain the solution to this problem, we estimate the corresponding first and second-order conditions and solve the corresponding system of equations. We then get the following
\begin{equation}
    x_{ij}^* = \frac{1}{2a}(u_{ij} - \frac{1}{m-1}\sum_{r\neq j} u_{ir}) \quad , \quad D_{ij}^* = \frac{m-1}{2} u_{ij} 
\end{equation}
\begin{equation}
    U_i(x_i^*,x_{-i}^*) = \underbrace{\sum_{j=1}^m P_{ij}^{(0)}u_{ij}}_{U_i(x_i=0,x_{-i}^*)} + \underbrace{\frac{p_i}{4a}\sum_{j=1}^m(m^2 u_{ij}^2 - (\sum_{s=1}^m u_{is})^2)}_{\geq 0} 
\end{equation}
Let's notice that the player possesses all the information needed to calculate the optimum deposit vector $D_i^*$. Also, the maximum utility achieved is composed by adding the expected value of not participating \footnote[4]{We can check that the expected value of not participating is equal to the expected value of participating with a null vote $x_i = (0,...,0)$, this is why we use the notation $u_i(x_i=0,x_{-i})$ as the utility of not participating.} and a non-negative term \footnote[5]{We can determinate that the term is non-negative by applying Cauchy-Schwartz. The equality only holds when $u_{ij}=u_{ir}$ for all $j,r = 1,...,m$, in other words, when the player has the same valuation for all alternatives.}.

The terms $R_{ij}^{(0)}$ and $R_{ij}^{(1)}$ present in the mechanism play a fundamental role. They guarantee the independence of the optimum vote vector $x_i^*$ from the belief's parameters $P_{ij}^{(0)}$ and $p_i$, which are unknown values to the mechanism.  
Since these parameters are different for each player, if $x_i$ would depend on them, the efficiency property would break.
The term $R_{ij}^{(1)}$ and the fact that the parameter $a$ is greater than zero, guarantees that the utility function is concave and a unique maximum exists. The variables $R_{ij}^{(1)}$ and $P_{ij}^{(1)}$ are proportional to $x_{ij}$. When $R_{ij}^{(1)}$ is multiplied by $(1-P_{ij}^{(1)})$ it generates a negative quadratic term.

\section{Properties}
The proposed mechanism recollects all necessary information and funds it needs to operate from the deposits made by participants and presents the following properties at equilibrium.    
%
\\~\\
\textbf{Non-indifferent players are incentiviced to participate}\\
An indifferent player has the same valuation for all alternatives, i.e., $u_{ij} = u_{ir}$ for $r=1,...,m$. If a player is not indifferent, then the utility obtained by voting optimally is strictly greater than not voting and implies that the mechanism incentivizes that player to participate, which is crucial for a mechanism to be implementable.
\begin{equation}
    (u_{ij})_{j=1}^m \neq (k,...,k) \quad \Longrightarrow \quad U_i(x_i^*,x_{-i}^*) > U_i(x_i=0,x_{-i}^*)
\end{equation}
We can check this through equation (9) since the equality of the second term holds only when the player has the same valuation for all alternatives, or in other words, the player is indifferent.
\\~\\
\textbf{Efficient}\\
We say a mechanism is \textit{efficient} if, given any realization of valuation vectors $(u_i)_{i=1}^n$, the alternative chosen by the mechanism is the one that maximizes the added valuations of players. So, given any realization of valuation vectors, we expect
\begin{equation}
    \argmax_{j} \sum_{i=1}^h x_{ij}^* = \argmax_{j} \sum_{i=1}^n u_{ij} 
\end{equation}
The reader can find a sketch of the proof in the appendix.
\\~\\
\textbf{SA-proof} \\
We say a mechanism is \textit{SA-proof} if every player participates only once at equilibrium. Suppose a player participates through $w$ different envelopes. Then, we always get (independently of the player's valuation vector) that the added utility over all participations of that player is less than or equal to the utility obtained by participating optimally only once. Which means  
\begin{equation}
    \sum_{q=1}^{w} U_q(x_q,x_{-i}^*) \leq U_i(x_{i}^*,x_{-i}^*)    
\end{equation}
where $x_{q}$ over $q=1,...,w$ are the resulting voting vectors of each participation made by the player. The intuition behind this result is that, by voting multiple times, the player generates a collective harm greater or equal to the individual benefit obtained by performing such action. Thus, the player's best response is limited to participating only once. The collective harm is reflected in the second term of equation (9) since it depends on the number of participants $h$ through the parameter $a$. This non-negative term of the utility function decreases when $h$ increases. The reader can find a sketch of the proof in the appendix.
\\~\\
\textbf{Non-negative Surplus}\\
All transfers made by the mechanism are non-negative, so no player could expect to get debt at the end of the voting process. Also, even though the mechanism is not balanced, it produces a non-negative surplus $S$. This surplus could finance projects of common interest. 
\begin{equation}
    R_i \geq 0  \quad , \quad S = \sum_{i=1}^h (D_i - R_i) \geq 0 
\end{equation}
Notice that a non-negative $S$ implies that the mechanism recollects all funds needed for the transfers since $\sum_{i=1}^h D_i \geq \sum_{i=1}^h R_i$.
\section{Conclusion}
We presented a voting mechanism that implements utilitarianism without requiring an identification or proof-of-personhood system over a scenario where risk-neutral players have valuations over each alternative and possess information over their valuations and beliefs (restricting to a family of priors distributions). Although the mechanism is not budget-balanced, the surplus is always non-negative and could finance common-interest projects. The mechanism recollects all necessary information and funds it needs from the participants' deposits.

\section{Appendix}

\textbf{Efficient:}
\\~\\
\textit{Sketch of proof:} 
\\~\\
Let's suppose each player participates only once, so $h=n$. We will demonstrate that for any configuration of vectors $(u_i)_{i=1}^n$ 
\begin{equation*}
    \argmax_{j} \sum_{i=1}^n x_{ij}^* = \argmax_{j} \sum_{i=1}^n u_{ij} 
\end{equation*} 
For each player $i$ and each alternative $j$ we have that
\begin{equation*}
    x_{ij}^* = \frac{1}{2a}(u_{ij} - \frac{1}{m-1}\sum_{r\neq j} u_{ir})
\end{equation*}
and
\begin{equation*}
    \sum_{i=1}^n x_{ij}^* = \sum_{i=1}^n \frac{1}{2a}(u_{ij} - \frac{1}{m-1}\sum_{r \neq j} u_{ir}) = \frac{1}{2a} \sum_{i=1}^n u_{ij} - \frac{1}{2a(m-1)}\sum_{i=1}^n \sum_{r \neq j} u_{ir}
\end{equation*}
Let's suppose that for some alternatives $j$ and $k$ 
\begin{equation*}
 \sum_{i=1}^n x_{ij}^* > \sum_{i=1}^n x_{ik}^*  
\end{equation*}
Then we have that
\begin{equation*}
 \frac{1}{2a} \sum_{i=1}^n u_{ij} - \frac{1}{2a(m-1)}\sum_{i=1}^n \sum_{r \neq j} u_{ir} > \frac{1}{2a} \sum_{i=1}^n u_{ik} - \frac{1}{2a(m-1)}\sum_{i=1}^n \sum_{r \neq k} u_{ir}  
\end{equation*}
\begin{equation*}
 \Longrightarrow \quad \sum_{i=1}^n u_{ij} - \frac{1}{m-1}\sum_{i=1}^n \sum_{r \neq j} u_{ir} > \sum_{i=1}^n u_{ik} - \frac{1}{m-1}\sum_{i=1}^n \sum_{r \neq k} u_{ir}  
\end{equation*}
\begin{equation*}
 \Longrightarrow \quad (m-1)\sum_{i=1}^n u_{ij} - \sum_{i=1}^n \sum_{r \neq j} u_{ir} > (m-1)\sum_{i=1}^n u_{ik} - \sum_{i=1}^n \sum_{r \neq k} u_{ir}  
\end{equation*}
\begin{equation*}
 \Longrightarrow \quad (m-1)\sum_{i=1}^n u_{ij} - \sum_{i=1}^n (\sum_{r \neq j,k} u_{ir} - u_{ik}) > (m-1)\sum_{i=1}^n u_{ik} - \sum_{i=1}^n (\sum_{r \neq j,k} u_{ir} - u_{ij})  
\end{equation*}
\begin{equation*}
 \Longrightarrow \quad (m-1)\sum_{i=1}^n u_{ij} - \sum_{i=1}^n \sum_{r \neq j,k} u_{ir} - \sum_{i=1}^n u_{ik} > (m-1)\sum_{i=1}^n u_{ik} - \sum_{i=1}^n \sum_{r \neq j,k} u_{ir} - \sum_{i=1}^n u_{ij}  
\end{equation*}
\begin{equation*}
 \Longrightarrow \quad (m-1)\sum_{i=1}^n u_{ij}  - \sum_{i=1}^n u_{ik} > (m-1)\sum_{i=1}^n u_{ik} - \sum_{i=1}^n u_{ij}  
\end{equation*}
\begin{equation*}
 \Longrightarrow \quad m\sum_{i=1}^n u_{ij} > m\sum_{i=1}^n u_{ik}
\end{equation*}
\begin{equation*}
 \Longrightarrow \quad \sum_{i=1}^n u_{ij} > \sum_{i=1}^n u_{ik}
\end{equation*}
Then, if an alternative adds more votes than any other it will also be the one that maximizes the added valuations of players over alternatives.
\\~\\
As an observation, any indifferent player would not affect the result if they decide to participate or not, even in the case they have positive valuations.
\\~\\
\textbf{SA-proof:} 
\\~\\
\textit{Sketch of proof:}
\\~\\
Suppose player $i$ wants to decide to participates $w$ times. We want to show that for any voting vectors $(x_q)_{q=1}^w$ we get
\begin{equation}
    \sum_{q=1}^{w} U_q(x_q,x_{-i}^*) \leq U_i(x_{i}^*,x_{-i}^*)    
\end{equation}
We can think of it as the player being $w$ different participants playing optimally with valuations that add up to the original valuation of player $i$ for each alternative $j$ (i.e., $u_{ij}$). Also they would share beliefs and know the other $w$ participants inputs and decide as a cartel.
\begin{equation}
    u_{ij} = \sum_{q=1}^{w} u_{qj}
\end{equation}
We then calculate the maximum utility that player $i$ can achieve for a particular configuration of vectors $(u_{qj})_{q=1}^w$ for each $j=1,...,m$, which is
\begin{equation}
    \sum_{q=1}^w U_q(x_q,x_{-i}^*) =  \sum_{j=1}^m P_{ij}^{(0)}u_{ij} + \sum_{q=1}^w \frac{p_i}{4a_{n+w-1}}\sum_{j=1}^m(m^2 u_{qj}^2 - (\sum_{s=1}^m u_{qs})^2)
\end{equation}
Where $a_{n+w-1}=(\frac{3}{2})^{n+w-1}$ is a parameter depending on $h=n+w-1$ the number of participants (we assume for notation simplification that other players participated once, however, it's not needed for the proof). We had from equation (9) that
\begin{equation}
    U_i(x_i^*,x_{-i}^*) = \sum_{j=1}^m P_{ij}^{(0)}u_{ij} + \frac{p_i}{4a_n}\sum_{j=1}^m(m^2 u_{ij}^2 - (\sum_{s=1}^m u_{is})^2) 
\end{equation}
Combining equations (16) and (17) we have that equation (14) holds (for any configuration of vectors $(u_{qj})_{q=1}^w$ such that satisfies equality (15) for all $j=1,...,m$) if
\begin{equation*}
    m^2\sum_{j=1}^m(\sum_{q=1}^w u_{qj})^2 - (\sum_{q=1}^w\sum_{j=1}^m u_{qj})^2 \geq \frac{a_n}{a_{n+w-1}} (m^2\sum_{q=1}^w \sum_{j=1}^m u_{qj}^2 - \sum_{q=1}^w(\sum_{j=1}^m u_{qj})^2)
\end{equation*}
which holds for $\frac{a_n}{a_{n+w-1}}\leq \frac{2}{3}$. 
\\~\\
Since
\begin{equation*}
   \frac{a_n}{a_{n+w-1}} = (\frac{2}{3})^{w} \leq \frac{2}{3} 
\end{equation*}
then we have that the inequality (18) holds for all $w \geq 1$.

\end{document}